\documentclass[12pt]{article}
\usepackage{epsfig}

\makeatletter
\@addtoreset{equation}{section}

\def\baselinestretch{1.2}
\def\href#1#2{#2}

\newcommand{\be}{\begin{equation}}
\newcommand{\ee}{\end{equation}}
\newcommand{\beq}{\begin{eqnarray}}
\newcommand{\eeq}{\end{eqnarray}}

\begin{document}
\begin{titlepage}

\begin{flushright}
IASSNS-HEP-00/85\\
hep-th/0012093\\
\end{flushright}

\vfil\vfil

\begin{center}

{\Large{\bf Traveling Faster than the Speed of Light in\\ \vspace{2mm}
 Non-Commutative Geometry} 
}

\vfil

\vspace{5mm}

Akikazu Hashimoto$^a$  and N. Itzhaki$^b$\\

\vspace{10mm}

$^a$Institute for  Advanced Study\\ School of Natural Sciences\\
Einstein Drive, Princeton, NJ 08540\\
aki@ias.edu\\

\vspace{10mm}

$^b$Department of Physics\\ University of California,
Santa Barbara, CA 93106\\
sunny@physics.ucsb.edu\\

\vfil
\end{center}

\begin{abstract}
\noindent We study various dynamical aspects of solitons in non-commutative
gauge theories and find surprising results. Among them is the
observation that the solitons can travel faster than the speed of
light for arbitrarily long distances.
\end{abstract}

\vspace{0.5in}

\end{titlepage}
\renewcommand{\baselinestretch}{1.05}  

\section{Introduction}

The study of solution to the classical equation of motion of field
theories on non-commutative geometries has led to the discovery of new
class of soliton solutions \cite{Gopakumar:2000zd} which have masses
of the order of the inverse of the non-commutativity scale.  In the
commutative limit, these solitons become infinitely massive and
decouple from the theory. In the case of gauge theories, these new
soliton solutions are localized magnetic vortices
\cite{Polychronakos:2000zm,Bak:2000ac,Aganagic:2000mh,Harvey:2000jb,Gross:2000ss}. One
remarkable aspect of these vortex solutions is the fact that they can
be treated exactly and explicitly by exploiting the algebraic
properties of gauge transformations. By now, there are systematic
accounts of the solution generating technique and the study of the
small fluctuations of these vortices in non-commutative gauge theories
in the literature
\cite{Polychronakos:2000zm,Bak:2000ac,Aganagic:2000mh,Harvey:2000jb,Gross:2000ss}.

In this article, we investigate the dynamics of these vortex solutions
beyond the small fluctuation approximation. Specifically, we
investigate the collective dynamics and the quantum effects of
magnetic vortices in 3+1 dimensional NCYM with space-space
non-commutativity.  To address the nature of the collective dynamics,
we construct the generalization of the vortex solutions to solutions
with finite velocity along the non-commutative coordinates. The
generalization is simple, but should not be dismissed as a trivial
boost of the static solution as in the case of solitons in ordinary
field theories.  In non-commutative theories, boosting along the
non-commutative direction is not a symmetry since it acts
non-trivially on the non-commutativity parameter $\theta_{\mu
\nu}$. Nonetheless, we will find an explicit solution describing a
vortex moving at some finite velocity.  Moreover, we construct
solutions which describe vortices moving with respect to each other.
We find that these vortices travel at velocities which can exceed the
speed of light.

To study the effect of quantum corrections, we exploit two independent
techniques: perturbation theory and the supergravity dual
\cite{Hashimoto:1999ut,Maldacena:1999mh}. In formulating the problem
of calculating the effect of integrating out some of the degrees of
freedom perturbatively at one-loop, we encounter an extremely useful
observation: this calculation is equivalent verbatim to some of the
calculations done earlier in the context of scattering gravitons and
membranes in Matrix theory \cite{Aharony:1997bh,Lifschytz:1997rw}. By
making simple change in the notation, we are able to read off the
results of these authors to draw interesting conclusions about the
quantum effects on the dynamics of non-commutative vortices. In
particular, some of the flat directions in the parameter space of
classical non-commutative vortices are lifted by the quantum effects.

The description of non-commutative gauge theories in terms of the
supergravity dual can be used to study the dynamics at very large 't
Hooft parameter where the quantum effects dominate.  The
non-commutative vortices correspond to D-string probes in the dual
supergravity background.  Therefore, the effective dynamics of the
non-commutative vortices is captured by the DBI action.  We conclude
also using this formalism that these vortices can move faster than the
speed of light.

The D-string probe in this supergravity background behave in many ways
like the ``long string'' in $AdS_3$
\cite{Maldacena:1998uz,Seiberg:1999xz,Maldacena:2000hw}. The D-string
probe feels a potential which becomes flat near the boundary at $U =
\infty$. This potential is consistent with the quantum corrected
potential on the moduli space computed perturbatively. The value of
the potential at $U =\infty$ is the same as the mass of the classical
solution.  Just as in the case of the long strings in anti de-Sitter
spaces, we conclude that the theory contains a continuum of states
above this gap.  We also discuss the implications of that conclusion
on the different phases of non-commutative gauge theories.

After the completion of this paper, we realized the fact that a
non-commutative vortex can travel faster than the speed of light was
also noted in \cite{Bak:2000im}.  Unlike \cite{Bak:2000im}, however,
we take the point of view that this phenomenon is real, and provide its
explanation in terms of string theory.

\section{Non-Commutative Vortex Strings}

In this section we describe the construction of magnetic strings in
non-commutative gauge theories in four dimensions.  To be specific, we
will concentrate on ${\cal N}=4$ supersymmetric NCSYM  which has
six adjoint scalars in addition to the gauge fields. Non-commutative
coordinates are taken to be along the $(x_2,x_3)$-plane. We will begin
by reviewing the construction of the static solutions. These solutions
can easily be generalized to the constant velocity solutions.

\subsection{Static solutions}

To describe the non-perturbative solution of non-commutative gauge
theories, it is more convenient to work in the operator formalism.
Following the notation of \cite{Aganagic:2000mh}, let us define the
complex coordinates
\be
z=\frac{1}{\sqrt{2}}(x^2+i x^3), \qquad {\rm such\ that}\qquad  [z,\bar{z}]=\theta.
\ee
We will compactify the $x_1$ direction on a circle. Then, considering
configurations which do not depend on $x_1$ and working in the
temporal gauge $A_0=0$, the action takes the form
\be
S=\frac{2\pi\theta L}{g_{YM}^2}\int dx_0  \mbox{Tr} \left(
-\partial_t \bar{C} \partial_t C +([C,\bar{C}]+1/\theta)^2 \right),\ee
where $C=-A_z+a^{\dag}$ and $L$ is the period of compact circle in the
$x_1$ direction.

The equations of motion are
\be\label{l}
\partial_t ^2 C = [C,[C,\bar{C}]],
\ee
while gauge fixing constraint yields
\be\label{ll}
[C,\partial_t \bar{C}]+[\bar{C},\partial_t C]=0.
\ee
Unlike ordinary gauge theories which do not admit static solutions
with non zero magnetic fluxes, there are such solutions to eq.(\ref{l})
\be\label{m}
C=(S^{\dag})^M a^{\dag}S^M +\frac{1}{\theta}
\sum_{i=0}^{M-1} l^i |i\rangle\langle i|,
\ee
where $l^i$'s are arbitrary complex numbers which correspond to the
position of the monopoles on the $(x_2, x_3)$-plane and $S$ is the shift
operator
\be S =\sum_{i=0}^{\infty} |i+1 \rangle \langle i |\ . \ee
The field strength  in the $z$ plane is 
\be\label{o}
\theta F=[C,\bar{C}]+1=\sum_{i=0}^{M-1} |i\rangle\langle i|, 
\ee
which implies that the total flux is
\be 
\frac{1}{2\pi}\int dx_2 dx_3\, F_{23} =\theta\,  \mbox{Tr} F =M.  
\ee 
It is interesting to note that eq.(\ref{o}) does not depend at all on
$l_i$.  At first sight this looks like a contradiction with the
statement that $l_i$'s are the locations of the strings in the
$(x_2,x_3)$-plane since the magnetic field clearly should depend on
the location of the strings.  However, in non-commutative gauge
theories, $F$ does not have a gauge invariant meaning and thus cannot
be used to specify the location of the strings.  In
\cite{Gross:2000ba} it was shown that the proper gauge invariant
generalization of $F$ is $F$ attached to the open Wilson lines of
\cite{Ishibashi:1999hs}.  By probing the solitons (\ref{m}) with these
operators, one finds indeed that $l_i$'s are the locations of the
strings \cite{Gross:2000ss}.  An alternative way to reach the same
conclusion is to consider the masses of the small fluctuations
\cite{Aganagic:2000mh}.

The energy of the solution (\ref{m}) is
\be\label{r}
E=\frac{2\pi\theta}{2g_{YM}^2}\mbox{Tr} F_0^2= \frac{\pi L M}{g_{YM}^2\theta}.
\ee
Notice that the total energy does not depend on $l_i$.  This means
that classically there are no static forces between the strings even
though they are non-BPS objects.

The ${\cal N}=4$ theory also contains six scalars in the adjoint which
enlarge the classical moduli space \cite{Aganagic:2000mh}.  Adding
\be
\varphi^{a}=\sum_{i=0}^{M-1} \varphi^{a}_{i} |i\rangle\langle i|
\ee
does not affect the classical equation of motion, total flux, or
energy.  However, it does have one very crucial effect. If the
expectation value is large enough, the solution becomes classically
stable.  To be more precise, in the absence of scalar expectation
values, there is a tachyon coming from the ``1-3'' string sector whose
mass is $m^2=-1/\theta$ \cite{Aganagic:2000mh,Gross:2000ss}.  In the
presence of scalars, the mass gets shifted \cite{Aganagic:2000mh}
\be\label{12} m^2=-\frac{1}{\theta}+\varphi^2.  \ee
Therefore, we see that for sufficiently large expectation values, the
solutions are {\em classically} stable. In fact, in the limit $\varphi
\rightarrow \infty$, all of the ``1-3'' strings become infinitely
massive, effectively decoupling the ``1-1'' sector from the ``3-3''
sector. In this limit, the effective action of the ``1-1'' sector
becomes simply that of $U(M)$ SYM in 1+1 dimensions.

\subsection{Constant velocity solutions}

Let us now describe a very simple, but yet interesting, generalization
to the static solutions.  Consider the same solution (\ref{m}), but
instead of setting $l_i$ to be a constant, take
\be
l_i=l^0_i+v_i t.
\ee
One can easily check that both the equation of motion (\ref{l}) and
the Gauss law constraint (\ref{ll}) are satisfied by this generalization.
The total magnetic flux is, of course, intact while the total energy
gets a kinetic energy contribution coming from the first term in the
action
\beq
 E &=&\frac{2\pi\theta}{g_{YM}^2}\int dx_0 \int_0^L dx_1 \mbox{Tr}
\left( \partial_t \bar{C} \partial_t C +([C,\bar{C}]+1/\theta)^2
\right) = \sum_{i=0}^{M-1} \left(m +  m v_i \bar{v}_i \right),
\nonumber \\
p & =& \frac{2\pi\theta}{g_{YM}^2}\int dx_0 \int_0^L dx_1 \mbox{Tr}
\left( \partial_t C \,   ([C,\bar{C}]+1/\theta)
\right) = \sum_{i=0}^{M-1} m v_i~, \label{pp} \\
m&=&\frac{\pi L}{g_{YM}^2 \theta}. \nonumber
\eeq
These equations lead to several remarkable observations about the
non-commutative solitons.  First, we see that not only are there no
static forces between the strings, there are no forces which depend on
the velocities of the strings either. Therefore, at least classically,
the vortices propagate like free particles.  Second, and more
importantly, the kinematics of the vortices are {\it
non-relativistic}.\footnote{This statement is closely related to the
observations regarding the non-relativistic dispersion relations of
closed strings in NCOS \cite{Klebanov:2000pp}.  See also
\cite{Gomis:2000bd,Danielsson:2000gi}.}  We will discuss this point
further in the concluding section.

\section{Quantum Effective Dynamics}

In the previous section, we saw that the non-commutative magnetic
solitons admit different surprising properties.  The aim of this
section is to see how much of this survives the quantum corrections.
The two methods at our disposal are perturbation theory which is valid
at small coupling and the supergravity description which is valid at
large coupling.

\subsection{Perturbation theory}

For simplicity, we consider only one unit of magnetic flux.  In that
case, the interesting part of the moduli space is parameterized by
$\varphi^a$.  As was briefly reviewed in the previous section, the
form of the potential of small fluctuations about the string solution
is \be V=T^2(-\frac{1}{\theta}+\varphi^2), \ee where $T$ represents
fluctuations coming from the ``1-3'' strings.  What we would like to
do is to integrate out $T$ to induce an effective potential for
$\varphi$.  This cannot be done for small $\varphi$ due to the
presence of the tachyon.  Therefore, we will focus our attention on
finding the effective potential for $\varphi^2 > 1/\theta$.

Fortunately, this somewhat tedious calculation was done earlier in the
context of scattering gravitons off membranes in Matrix theory
\cite{Aharony:1997bh,Lifschytz:1997rw}. Modulo the trivial
modification between the 0-brane soliton in 2+1 dimensions and the
1-brane soliton in 3+1 dimensions, the calculation
\cite{Aharony:1997bh,Lifschytz:1997rw} is verbatim the calculation we
wish to do here. All that we need to do is to appropriately smear
equation (66) of \cite{Lifschytz:1997rw}
\be
V = {3 N \over 16}  {c^3 \over b^5} \Longrightarrow V = {L N \over 8 \pi} {c^3 \over b^4}
\ee
($N$ is the rank of the gauge group) and to appropriately relabel the
variables: $c = {1/ \theta},~ b = \varphi_0,$ to find that the
effective potential as a function of $\varphi_0$ is
\be\label{lo} 
V_{eff}=-\frac{(2\pi)^3 LN}{4 (2 \pi \varphi_0)^4 \theta^3}.
\ee
We see, therefore, that the classical moduli space is lifted by
quantum corrections which tend to lower the expectation value of the
scalars.  If one starts with the vortex at some large but finite value
of $|\varphi_0|$, the quantum correction will lift the flat direction
and the expectation value of $|\varphi_0|$ will start rolling
down. Eventually, $|\varphi_0|$ becomes small enough and some of the
``1-3'' strings become tachyonic, where we must abandon the
perturbative calculation all together.

From eq.(\ref{lo}) one finds that the lifetime of the soliton is
roughly $\varphi_0^3 \theta^2 /\sqrt{\lambda}$, which means that even
quantum mechanically, the solitons can live for an arbitrarily long
time (by taking $\varphi_0$ to be large).

\subsection{Supergravity description}

The relevant parts of the supergravity background are given by
\cite{Hashimoto:1999ut,Maldacena:1999mh}
\beq\label{bg}
&& \frac{ds^2}{\alpha'} =   {U^2 \over \sqrt{\lambda}}(- dx_0^2 + dx_1^2) +
 {\sqrt{\lambda} U^2 \over \lambda + \Delta^4 U^4} (dx_2^2 + dx_3^2) +
 {\sqrt{\lambda} \over U^2} dU^2 + \sqrt{\lambda} d \Omega_5^2 
 \\ \nonumber
&&e^{\phi}  = { g_{YM}^2 \over 2 \pi} \sqrt{{\lambda \over \lambda + 
\Delta^4 U^4}},~~~~~
B_{23}  =   - {\alpha' \Delta^2 U^4 \over \lambda + \Delta^4 U^4},~~~~~
A_{01}  =  {2 \pi \over g_{YM}^2} {\alpha' \Delta^2 U^4 \over  \lambda} 
\eeq
where $\lambda = 2 g_{YM}^2 N$ and $\theta_{23} = \theta = 2 \pi \Delta^2.$  

\begin{figure}
\centerline{\psfig{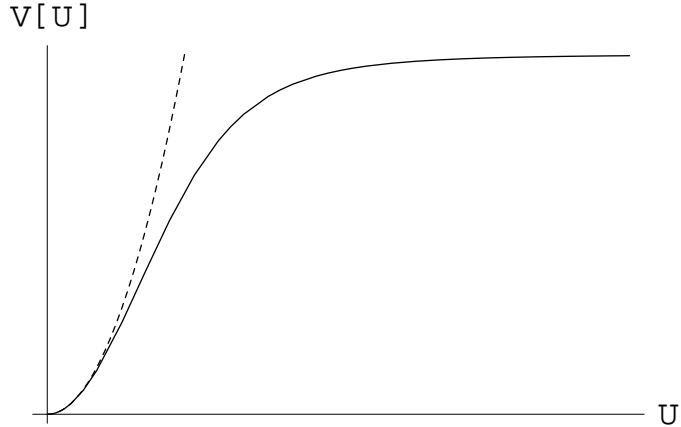}}
\caption{
The D1-branes potential as a function of the radial coordinate. The
solid line is the potential in the non-commutative case in which with
a finite amount of energy the D1-brane can reach the boundary.  The
dotted line is the potential in the commutative case. \label{figa}
}
\end{figure}

The dual of the magnetic string considered in the previous sections is
a D1-brane oriented along $x_0$ and $x_1$.  The action of such a
D-string in the background (\ref{bg}) is 
\be V(U) = {L \over g_{YM}^2} \left( {U^2 \sqrt{\lambda + \Delta^4 U^4}
 \over \lambda} - {\Delta^2 U^4 \over \lambda} \right). \label{potential}
\ee
Simple inspection near $U=0$ gives
\be V(U) = {L U^2 \over g_{YM}^2 \sqrt{\lambda}}+{\cal O}(U^6)
 \label{smallU},\ee
whereas at infinity
\be V(\infty) = {L \over 2  g_{YM}^2 \Delta^2}. \label{gap}\ee
The important point is the fact that this potential is finite.  The
potential which starts out growing quadratically in $U$ becomes flat
at large $U$ and converge to (\ref{gap}). See figure \ref{figa} for an
illustration. Because the potential becomes flat in the $U \rightarrow
\infty$ limit, the brane sitting at $U = \infty$ is a solution to the
equations of motion.  Such a configuration of strings winding along
the boundary of space-time at finite cost in energy is known as the
``long string'' and has appeared in several contexts of AdS/CFT
correspondence
\cite{Maldacena:1998uz,Seiberg:1999xz,Maldacena:2000hw}.  Note that in
the commutative limit, the quadratic dependence (\ref{smallU})
persists for all values of $U$, and the mass of the would-be long
D-string becomes infinite.

The form of the potential illustrated in figure \ref{figa} suggests
that a D-string which starts in the neighborhood of $U=0$ can ``escape
to infinity'' if it carries enough kinetic energy. The minimum energy
required to escape to infinity is the gap (\ref{gap}).  At energies
below this gap, the D-string will be bounced back to the near horizon
region, whereas at energies above this gap, the D-string will escape
to infinity.  We emphasize, however, that it will take infinite time
for the brane to reach the boundary.  This statement is the same (up
to time reversal) as the statement of the previous section that the
lifetime of the soliton can be made arbitrarily long. The kinetic
energy of the string escaping to infinity is therefore not quantized,
giving rise to a continuum of states analogous to what was seen in the
case of $AdS_3$.

Note that the mass of the long string (\ref{gap}) is {\em exactly} the
same as the mass found in the free theory limit (\ref{r}).  This is
surprising since these solutions are non-BPS, and in general there is no
reason for their masses to be protected.\footnote{For example, in
\cite{Drukker:2000wx} a holographic description of some non-BPS branes
was found to receive correction to their masses due to strong
coupling.}  The reason why they are is the fact that at large $U$
(compared to $1/\Delta$) the supergravity solution looks like the near
horizon region of D1-branes oriented along $(x_0,x_1)$ and smeared
along the $(x_2,x_3)$ plane.  A D1 probe with the same orientation is
of course BPS and thus the potential at infinity is protected. We will
find more evidence for this claim in the discussion below.

In the previous section, we saw that the quantum corrections lifted
the potential by introducing the term (\ref{lo}). To see this term in
supergravity, we expand the exact supergravity potential for the
D1-brane around $U=\infty$ and find
\be
V(U) = {L \over 2 \Delta^2 g_{YM}^2} - {N L \over 4 \Delta^6 U^4} + 
{\cal O}(U^{-8})
\ee
in exact agreement with the field theory result (\ref{lo}) with the
usual identification \cite{Maldacena:1998kk} $U=|2 \pi \varphi_0|$.

While such a nice agreement with the weakly coupled field theory
results is found at large $U$, at small $U$ the strongly coupled
description implies a completely different physics.  Recall that in
perturbation theory, we encountered a tachyon at small $\varphi$ (or
small $U$) coming from the ``1-3'' strings.  For large 't Hooft
coupling, we do not encounter such a behavior at small values of
$U$. The background geometry simply becomes that of $AdS_5 \times
S_5$, and the D-string will simply continue to fall toward the horizon
in this background. In this region, the physics of the falling
D-string is identical to the one described in
\cite{Banks:1998dd,Balasubramanian:1998de}. From any finite $U$, it
takes infinite time for the D-string to reach the horizon. The falling
of the brane is to be interpreted as the spreading of the flux at
approximately the speed of light.

In $AdS_3$ the effective description of the long string is the
Liouville theory \cite{Seiberg:1999xz} which plays an important role
in the duality.  To find the effective description of the long string
in our case we consider fluctuations of the D1-brane:
\beq
F_{01} & = & \partial_0 A_1(t,x_1) - \partial_1 A_0(t,x_1),\nonumber
 \\ \nonumber x_2 &=& 2 \pi \Delta^2 \varphi_2(t,x_1), \\ x_3 &=& 2
\pi \Delta^2 \varphi_3(t,x_1),\\ \nonumber U &=& U_0 + 2 \pi
\varphi_4(t,x_1),\\ \nonumber \vec{\Omega}_5 & = & {2\pi\over U_0}
\varphi_i(t,x_1), \quad i= 5\ldots 10.  \nonumber \eeq 
Substituting this into the DBI action and taking the $U_0 \rightarrow
\infty$ limit leads to an effective action for the fluctuations given
by
\be S = {1 \over 2 \Delta^2 g_{YM}^2} - {(2 \pi \Delta)^2 \over
g_{YM}^2} \left( {1 \over 4 } F^2 + \sum_{i=2 \ldots 10} {1 \over 2}
(\partial \varphi_i)^2\right), \ee
which is precisely the bosonic part of the action of ordinary
supersymmetric Yang-Mills theory in 1+1 dimensions with 16
supercharges and the two dimensional coupling constant $g_{YM}^2/2
\pi \theta$.  It is worthwhile to emphasize that the quadratic form of
the action is not a result of taking the field strength to be small.
On the contrary, the DBI correction to the action is controlled by
$F^2/U_0^4$, and in the limit $U_0 \rightarrow\infty$ keeping $F$
fixed, the action becomes exactly SYM.  In other words, $1/U_0^2
\rightarrow 0$ replaces the role of $\alpha' \rightarrow 0$ of the
usual decoupling limit. This is the same as the $\varphi_0 \rightarrow
\infty$ limit which decouples the ``1-1'' sector from the ``3-3''
sector described in the previous subsection.

It is also very easy to see in the supergravity dual that the D1-brane can move
faster than the speed of light.  Consider the dependence of the
D1-brane action on the velocity $v^2=(\partial_0 x_2)^2+(\partial_0
x_3)^2$
\be\label{ji} S=\frac{\Delta^2 U_0^4}{\lambda g_{YM}^2}\left(
\sqrt{1+\frac{\lambda}{\Delta^4 U_0^4}} \sqrt{1-\frac{v^2}{1+\Delta^4
U_0^4/\lambda}}-1\right) .
\ee
We see that at large $U_0$ the bound on the velocity of the string, as
measured in the field theory units, is not $1$ (the speed of light)
but rather $U_0^2\theta$ which goes to infinity.  Moreover, for
$U_0\rightarrow\infty$, eq.(\ref{ji}) yields the non-relativistic
dispersion relation of (\ref{pp}).  It should be noted that the
velocity of the string as measured by a local observer in the
supergravity background is, of course, smaller than the speed of
light.

\section{Phases of NCSYM}

In the previous section, we saw that there are 1+1-dimensional ordinary
$U(M)$ SYM theories within 3+1 dimensional NCSYM above a mass gap
which is proportional to $M$.  This statement has important
implications on the phases\footnote{We use the word ``phase'' here in
the sense of \cite{Itzhaki:1998dd}. The phase boundaries in this
context may or may not be a phase transition in the sense of
non-analyticities in the thermodynamic observables.} of NCSYM. In this
section we explore this notion and make contact with
\cite{Gubser:2000mf}.

First let us recall the different phases of the theory at low energies
which we parameterize using the 't Hooft coupling and the S-dual 't
Hooft coupling, $\tilde{\lambda}=N^2/\lambda$.

\begin{tabular}{r l}
{\bf I} & $\lambda \ll 1$ is the weakly coupled NCSYM phase.\\
{\bf II} & $1 \ll \lambda$, $1 \ll \tilde \lambda$  is the dual supergravity phase.\\
{\bf III} &$  \tilde \lambda \ll 1$  is the  weakly coupled  phase of the S-dual theory.
\end{tabular}
 
These regions are non-overlapping. The intermediate region {\bf II}
exists only for large $N$.  The S-dual theory in the region {\bf III}
is the NCOS theory
\cite{Gopakumar:2000na,Seiberg:2000ms,Barbon:2000sg}.  Although NCOS
is a valid description at low energies, one finds a richer phase
structure in the theory at higher energies. The basic idea is similar
to what was discussed in \cite{Itzhaki:1998dd}, and was partially
discussed already in \cite{Harmark:2000wv}.

One way to see that the theory goes into a new phase at higher
energies is to note that the dual supergravity description remains
valid even in region {\bf III} for sufficiently high temperatures. The
reason is that in this supergravity background, the dilaton is $U$
dependent and goes to zero for large $U$. Note that this is very
different from the behavior of the commutative theory.  Among other
things, this means that the supergravity description is applicable in
the UV even for small values of $N$, {\it even for $U(1)$}, as long as
we take the coupling $\lambda$ to be large enough!  It is easy to
verify that the range of validity\footnote{A
supergravity description is valid when both the dilaton and the
curvature in string units are small.  Just like in the case of
$AdS_5\times S^5$, the curvature of the background (\ref{bg}) is of
the order of $1/\sqrt{\lambda}$ and is therefore small in region {\bf
III}.  One important distinction of the non-commutative case is the
fact that the dilaton is not constant but rather goes to zero at
infinity.  Hence, the supergravity description of NCSYM is valid even
in region {\bf III} for sufficiently large $U$. For smaller values of
$U$, the dilaton starts growing and eventually we have to apply
S-duality. In the S-dual background, the dilaton is getting smaller as
we go further in the IR, but the curvature starts growing and becomes
of order one at the boundary (\ref{sd}). This is similar to the phase
diagram of D5-branes \cite{Itzhaki:1998dd}.} of the supergravity description is
$U \gg \sqrt{\lambda / N \theta}$, or equivalently
\be\label{sd} T \approx {U \over
\sqrt{\lambda}} \gg \frac{1}{\sqrt{N} \Delta}.  \ee 
Note that since a theory is defined at the fundamental level in the UV
(where the number of degrees of freedom is maximal), what we are saying
is that at the fundamental level the definition of the theory is in
terms of the supergravity dual.  The NCOS phase is an effective
description valid {\it only} in the deep IR.

\begin{figure}
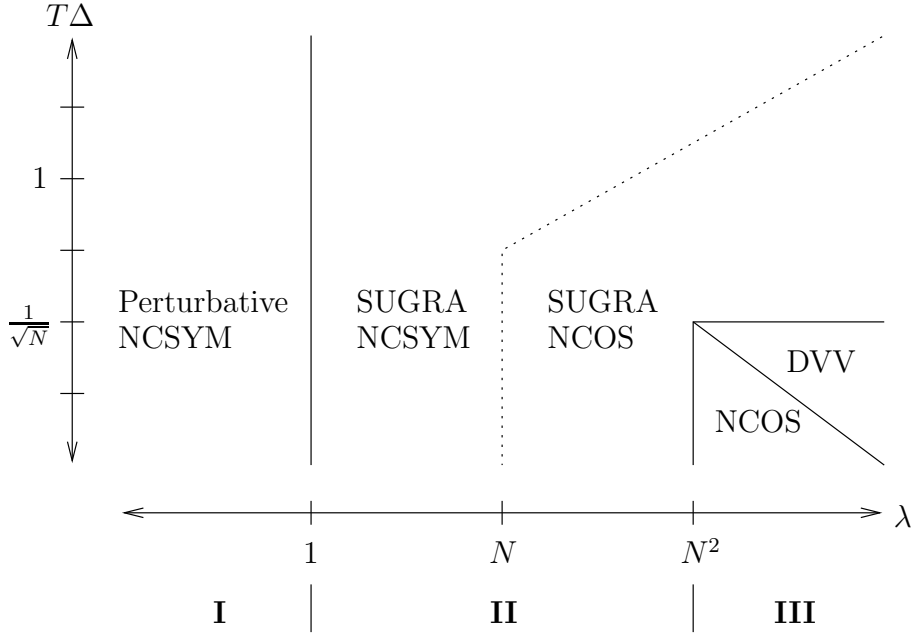

\centerline{\input phase.pstex_t}
\caption{Phase diagram of NCSYM.  The horizontal axis parameterizes the 't Hooft coupling and the vertical axis parameterizes the temperature. Both axes are logarithmic. \label{figb}}
\end{figure}

The NCOS theory in region {\bf III} undergoes a
Hagedorn phase transition \cite{Gubser:2000mf} at temperature 
\be\label{pop}
T_{H}=\frac{1}{2\pi \alpha_{eff}}=\frac{1}{g_{YM}\sqrt{\theta}}.
\ee
The perturbative NCOS description is valid below this temperature, but
above $T_H$, the system is better described as a thermal ensemble of
free strings. The physics of these strings resembles that of the DVV
strings \cite{Dijkgraaf:1997vv}. The complete picture of the phase
structure can be summarized in a phase diagram. See figure \ref{figb}.

The Hagedorn transition is a phenomenon of weakly coupled NCOS theory
in region {\bf III}.  When the NCOS coupling constant $\tilde \lambda$
is taken to be sufficiently large, one finds oneself in region {\bf
II}, where there is no Hagedorn transition.  Therefore, supergravity
is the proper description of the theory in region {\bf II} for all
energy scales.  We also expect to see that the liberated strings
re-condense at temperatures above the DVV/SUGRA phase boundary
(\ref{sd}) in region {\bf III}.  In other words, there should be no
overlap between the DVV and the SUGRA phases.  Let us examine this
assertion more closely.

First, let us recall the basic mechanism of \cite{Gubser:2000mf} which
drives the Hagedorn phase transition in region {\bf III}. NCOS is a
certain decoupling limit of D3-F1 bound state. There is an energy gap
to liberate the F1 from the D3. However, at temperatures above $T_H$,
the entropic contribution will compensate the free energy, making the
configuration of liberated F1 strings the preferred state of the
system.

In order to argue that the Hagedorn transition ceases to occur in
region {\bf II}, it suffices to show that one does not gain in free
energy by liberating the F1.  Note that the F1 in NCOS is dual to
the D1 probe in the supergravity description of NCSYM which we
considered in the previous section.  Therefore, what we have to do is
to compute the effect of the finite temperature on the mass gap
(\ref{gap}). This can be inferred from the action of the string probes
in the finite temperature generalization of the supergravity
background. The only relevant effect of the temperature\footnote{We
are not careful here with numerical factors of order $1$.} is to
modify the supergravity background by multiplying $g_{00}$ by a factor
$\left(1-\frac{U_0^4}{U^4}\right)$ where $U_0=T\sqrt{\lambda}$. This
will give rise to a new expression for the energy gap
\be
E_{gap}={L \over   g_{YM}^2 \Delta^2}+LT^4\theta N\ ,
\ee
which states that the energy gap increases as we increase the
temperature.  The negative contribution from the entropy to the free
energy remains the same as in \cite{Gubser:2000mf}. Therefore the free
energy per unit length of the string probe is
\be\label{lk}
f=\frac{F}{L}=\frac{E-ST}{L}={1 \over   g_{YM}^2 \Delta^2}+T^4\theta N -T^2.
\ee
The strings are liberated when $f\leq 0$.  One can easily see that
this is possible only in region {\bf III} and hence the DVV phase is
not realized in region {\bf II}.  Moreover, even in region {\bf III}
the DVV description is realized only for
\be
\frac{1}{\sqrt{\theta N}}>T>\frac{1}{g_{YM}\sqrt{\theta}},
\ee
which is in exact agreement with eqs.(\ref{sd}) and (\ref{pop}).\footnote{
Similar conclusion, using a different approach, was reached by J. Barbon and
 E. Rabinovici (to appear).}

\section{Conclusion}

In this article, we considered various dynamical aspects of solitons
in non-commutative gauge theories. The most dramatic and surprising
observation of this study is the fact that these solitons can travel
faster than the speed of light.  We will comment on this remarkable
result in the remainder of this paper.

Traveling faster than the speed of light is, of course, not possible
in a Lorentz invariant theory.  If Lorentz invariance is broken, one
might naively expect to be able to travel faster than the speed of
light for at most the distances of the order of the scale of broken
Lorentz symmetry. In nature, this distance is very small since Lorentz
symmetry is tested to hold to very small length scales.  The result we
obtain in this paper illustrates that this picture is too naive.
Non-commutative gauge theories with ${\cal N}=4$ supersymmetry breaks
Lorentz invariance only for length scales smaller than the
non-commutativity scale.  Yet, there is no bound on the distances the
non-commutative solitons can travel at speeds faster than the speed of
light.\footnote{A different mechanism for transmitting signals faster
than the speed of light, which relies on broken Lorentz invarance at
large distances as a result of UV/IR mixing \cite{Minwalla:1999px} in
a theory with less supersymmetry, was discussed in
\cite{Landsteiner:2000bw}.}  This is remarkable especially in light of
the fact that we have not made the time coordinate non-commutative.

\begin{figure}
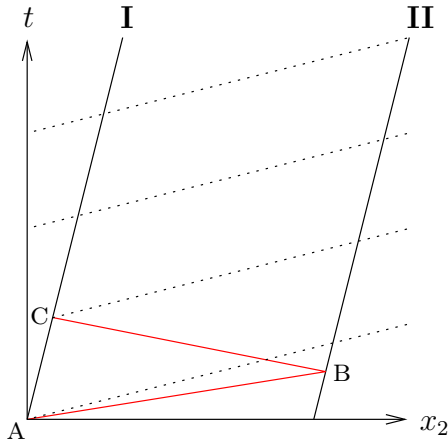

\centerline{\input causality.pstex_t}
\caption{
``Back to the Future.'' Two observers, {\bf I} and {\bf II}, are in motion
relative to the preferred frame $(t,x_2)$.  The dashed lines represent
the constant time slices in the frame of {\bf I} and {\bf II}.  {\bf
I} can send a signal from {\sc a} which {\bf II} receives at a point
{\sc b} in the past.  However, when {\bf II} sends the information back to
{\bf I} it is at the point {\sc c} which is always in the future of
{\sc a} as seen by {\bf I}.}
\label{fig8}
\end{figure}

One might worry that being able to travel faster than the speed of
light will cause causality to break down. Indeed, a signal traveling
faster than the speed of light looks to a moving observer like a
signal traveling backwards in time.  However, problems with causality
arise only if an observer can send a signal from the future to itself
in the past.  This turns out not to be possible, essentially because
there is a preferred frame in which the information can travel
arbitrarily faster than the speed of light, but not backwards in time.
This is the frame of the stationary observer. Any signal transmitted
by an observer can only be received by the same observer in the
future.  See figure \ref{fig8} for more details.

Along similar lines, one can see that problems with unitarity do not
arise. Non-commutative geometry as seen by a moving observer is
certainly a strange world. The action has infinitely many time
derivatives, and a Hamiltonian cannot be defined in general. However,
precisely because there is a preferred frame in which
non-commutativity is strictly space-like, a Hamiltonian {\it can} be
defined in that frame. In other words, the criterion of
\cite{Gomis:2000zz} is not violated by theories with only space-like
non-commutativity.

While propagation at speeds faster than the speed of light may seem
strange from the conventional field theory points of view, its origin
can be understood in very simple terms in the language of the
underlying string theory. The ``speed of light'' is defined as the
speed at which the ``3-3'' strings propagate in the {\it open string}
metric since the large $B$-field along the D3-brane affects their
dynamics. The non-commutative solitons, on the other hand, are
D1-branes. The large $B$-field is transverse to the D1-brane, and
consequently does not affect the dynamics of the ``1-1'' strings.
Therefore, the D1-branes are living in the {\it closed string} metric.
As was explained in \cite{Douglas:1998fm,Seiberg:1999vs}, the ratio
between the closed string metric and the open string metric goes to
zero in the field theory limit.  Therefore, the relativistic
dispersion relation in the closed string metric becomes the
non-relativistic dispersion relation in the open string coordinates in
this limit. This should also have important implications for coupling
non-commutative field theories with gravity.

Conventional ideas for traveling faster than the speed of light
involve ``short-cuts'' in space-time such as the wormholes and the
warps. The result presented in this paper offers a new alternative:
non-commutatizing the universe. Or better yet, nature may already be
non-commutative at small length scales.  One can imagine a scenario
where one is living on a brane with a large background $B$-field. One
can then transmit signals at speeds faster than the speed of light by
sending it in the bulk (closed string) metric.  It should be amusing
to study this further.

\section*{Acknowledgement} 

We would like to thank
S.~Giddings,
D.~Gross,
I.~Klebanov,
D.~Kutasov,
J.~Maldacena, 
N.~Seiberg,  
M.~Srednicki, and
E.~Witten
for discussions. Part of this work was carried out while
we were visitors at the University of Chicago, and we would like to
thank the members of the theory group for the warm hospitality. The
work of AH is supported in part by the Marvin L.~Goldberger fellowship
and the DOE under grant No.\ DE-FG02-90ER40542. The work of NI is
supported by the NSF under grant No.\ PHY97-22022.

\bibliography{soliton} \bibliographystyle{utphys} 

\end{document}